\documentclass[11pt,preprint]{aastex}

\def\lsim{\mathrel{\hbox{\rlap{\lower.55ex \hbox {$\sim$}}\kern-.0em 
\raise.4ex \hbox{$<$}}}}  
\def\gsim{\mathrel{\hbox{\rlap{\lower.55ex \hbox {$\sim$}}\kern-.0em 
\raise.4ex \hbox{$>$}}}}

\def\farcs{\hbox{$.\!\!^{\prime\prime}$}}

\def\arcsec{\hbox{$^{\prime\prime}$}}

\begin{document} 
\title{Keck Spectroscopy and HST Imaging of GRB~000926: Probing a Host Galaxy at $z = 2.038$} 
\author{S.~Castro\altaffilmark{1,3},  
T.~J.~Galama\altaffilmark{1},  
F.~A.~Harrison\altaffilmark{1}, 
J.~A.~Holtzman\altaffilmark{2},   
J.~S.~Bloom\altaffilmark{1}, 
S.~G.~Djorgovski\altaffilmark{1}, 
S.~R.~Kulkarni\altaffilmark{1}
} 
 
\begin{abstract} 
We present 
early-time Keck spectroscopic observations and
late-time Hubble Space Telescope (HST) imaging of GRB\,000926.  The HST
images show a small offset between the optical transient and the
compact host galaxy. Combined with the large equivalent widths
measured for metallic absorption lines by the Keck {\it Echellette
Spectrograph and Imager} (ESI) and the {\it Low-Resolution Imaging
Spectrometer} (LRIS), this indicates that the GRB exploded near the
center of its host.  The ESI spectroscopy reveals two
absorption systems centered at $z = 2.0379 \pm 0.0008$ with a velocity
separation of 168~km~s$^{-1}$, which we interpret as being due to
individual clouds in the host galaxy.  The ratios of chromium to zinc
equivalent widths indicates the host is depleted in dust relative to
local values to a similar
degree as damped Lyman alpha systems at the same redshift.  Further,
the two clouds appear to have similar relative metal abundance and dust to gas
ratio.  If one cloud is associated with the GRB site, this implies the
explosion did not significantly alter the surrounding environment.
\end{abstract} 
%\submitted{Draft version \today} 
 
\altaffiltext{1}{Palomar Observatory, 105-24, California Institute of 
Technology, Pasadena, CA, 91125.} 
\altaffiltext{2}{Department of Astronomy, New Mexico State University, Box 
30001, Department 4500, Las Cruces, NM, 88003-8001.} 
\altaffiltext{3}{Infrared Processing and Analysis Center 100-22, California 
Institute of Technology, Pasadena, CA, 91125.}

\section{Introduction} 
 
Observational evidence indicates that gamma-ray bursts explode in
luminous regions of galaxies, suggesting the progenitors of these
events are a population associated with galactic disks, and may
descend from massive stars.  Hubble Space Telescope (HST) imaging
reveals that GRB optical transients have smaller offsets relative to
the centers of their host galaxies than would be expected for
a halo population \citep{bdk01}.   A few events have exhibited large optical
extinctions, such that they are visible only at X-ray, IR and radio
wavelengths \citep{ggt+97,tfk+98,dfk+01}. This indicates these exploded in
or behind dense regions of the host galaxy.  The large equivalent
widths of metallic absorption lines seen in optical spectra of 
optical transients also point to a location within the host
disk.

As a disk population, a typical line of sight towards a GRB will
intersect a significant column of the galactic insterstellar medium.
Intermediate and high-resolution absorption spectroscopy can therefore
be used to probe the metal content of the host, as well as individual
structures, if they are separated by sufficient doppler shifts. Ratios
of metal lines can also be used to probe the dust content.  For
example, the Zn~II ($\lambda$2025, 2062) doublet is a good indicator
of metallicity, as it has near-solar abundance in gas with little
depletion onto interstellar grains \citep{pbh90}.  Cr, on the other
hand, exists mostly in interstellar grains, and is among the most
heavily depleted elements in the gas phase of the ISM.  Measuring both
the Zn~II doublet as well as the Cr~II triplet ($\lambda$2055, 2061,
2065) therefore provides a qualitative measure of the dust to gas
ratio \citep{pbh90,psh+94}.
 
Comparing chemical composition and dust content in a sample of GRB 
host galaxies to, for example, damped Ly$\alpha$ systems at similar 
redshift would indicate whether GRBs occur in regions with unusual 
metal enrichment.  Since it is unclear what factors, other than mass, 
play a role in determining which stellar progenitors produce GRBs, 
this could provide direct clues to the explosion.  It is also 
interesting to look for anomalously low dust to gas ratios.  \citet{wd00} 
and \citet{fkr01} 
have theorized that the early, hard radiation from the GRB and its 
afterglow should destroy dust in the circumburst environment.  The 
destruction of dust grains would release metals, and increase the 
relative equivalent widths of those elements preferentially condensed 
on dust grains.  
 
In this paper, we present early-time spectroscopic observations with the 
W.M. Keck of the afterglow of GRB\,000926 using the {\it Echellette 
Spectrograph and Imager} (ESI) and the {\it Low-Resolution Imaging 
Spectrometer} (LRIS), combined with imaging of the host galaxy
by HST.     
 
\subsection{GRB\,000926} 
 
The {\it Inter-Planetary Network} (Ulysses, Konus-Wind, and Near)
detected the long-duration ($t_{\gamma} = 25$~s) event GRB\,000926 on
2000 Sep 26.993 UT \citep{hmg+00}.  \citet{gcc+00} and \citet{dfp+00}
identified the afterglow.  The discovery and optical lightcurves are
presented in \citet{fgd+01}.  Spectra of the
afterglow from the Nordic Optical Telescope yielded an absorption
redshift of 2.066 \citep{fmg+00}, later refined to 2.0379 $\pm $
0.0008 from Keck spectroscopy \citep{cdk+00}.
 
Ground-based multi color lightcurves of the afterglow show a 
steepening beginning $\sim$ 1.2 days after the event \citep{phg+01}, 
providing evidence that the ejecta are collimated into a jet.  The 
gamma-ray energy release, corrected for the implied collimation, is 
$2.2 \times 10^{51}$~erg. A fit of the data to an afterglow model 
requires modest extinction in excess of the Galactic value along the 
line of sight (A$_V$ in the range 0.11 -- 0.82 mag; see 
\citet{phg+01}), presumably due to the host galaxy. Synthesis of the 
broad-band data set resulting from {\it Chandra} X-ray observations, 
continued optical monitoring by HST and 
{\it Very Large Array} radio observations reveal evidence that the 
cooling is dominated by Inverse Compton scattering, with the IC 
component directly observable \citep{hys+01}.  This implies that the 
GRB exploded in a moderately dense, $n \sim 30$ cm$^{-3}$ medium, 
consistent with a diffuse interstellar cloud environment. 
 
\section{Observations} 
 
\subsection{Moderate Resolution Spectroscopy}

Starting at about UT 2000 September 29.26 L. Cowie obtained two
spectra, each of 1800-s duration, using ESI \citep{em98} in the
Echelle mode. In this mode, the spectrum covers the range 3900
$\rm\AA$ to 10900 $\rm\AA$ over ten orders. The native spectral
resolution of the instrument is 11.4 km s$^{-1}$ pixel$^{-1}$. The two
spectra were obtained at an airmass of about 1.33 and position angle
of 128$^\circ$, about 15$^\circ$ from the parallactic angle.
 
We employed two procedures for data reduction; one using standard IRAF 
packages and the other using the {\it Mauna Kea Echelle Extraction} - 
MAKEE program (written by T. Barlow).  The two reductions agreed to 
within statistical errors.  We reduced the echelle orders using a 
bright star to trace each individual order prior to extraction. Each 
exposure was optimally extracted and background subtracted. We 
identified lines in CuAr lamp spectra, individually extracted using 
the object's apertures, and determined the wavelength scale by 
polynomial fitting the line positions using a mean r.m.s.  of 0.09 
$\rm\AA$. The two spectra were then added to yield the final spectrum.

\begin{figure}[] 
\epsscale{0.75}
\plotone{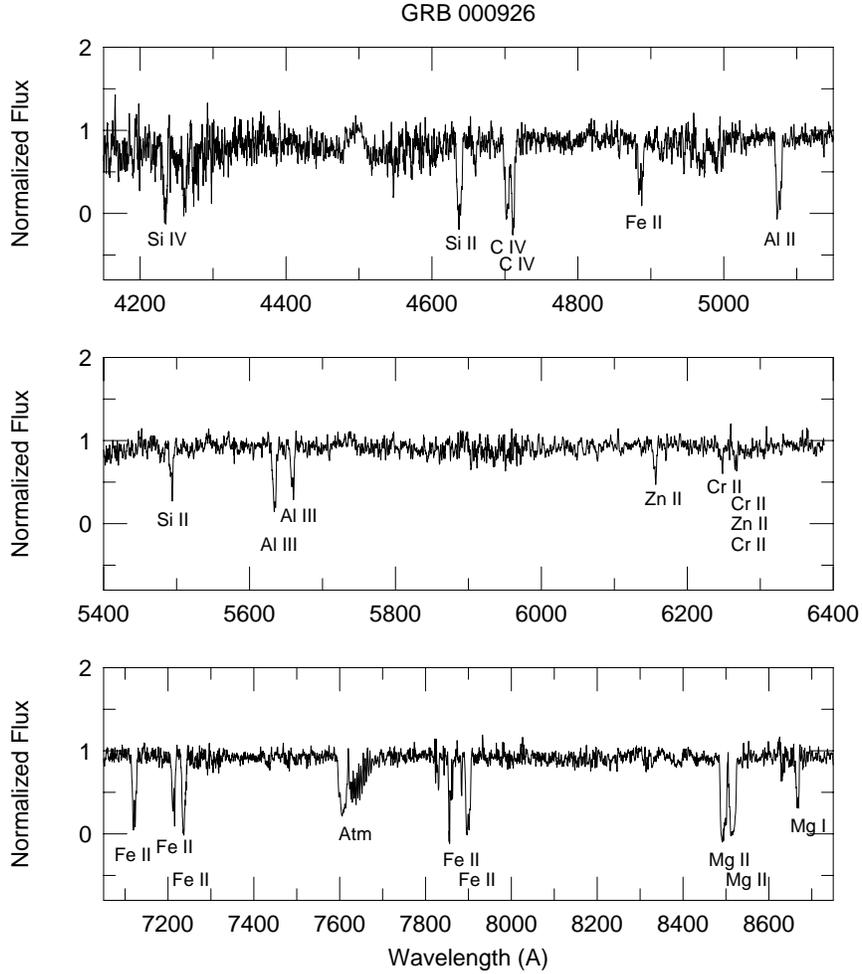}
\caption{The ESI spectrum of GRB\,000926 on UT 2000 September 
29.26. Indicated are the line identifications corresponding to a 
redshift $z = 2.0372 \pm 0.0008$. The absorption feature around 7600 
$\rm\AA$ is an atmospheric (Atm) line due to telluric O$_2$. The spectrum 
has been smoothed by a boxcar of 7 pixels, equivalent to 1.12 
$\rm\AA$. } 
\label{fig:esispec} 
\end{figure} 
 
\subsection{Low Resolution Spectroscopy} 
 
Low resolution spectra of the GRB 000926 afterglow were obtained using
the {\it Low Resolution Imaging Spectrometer} (LRIS) \citep{occ+95}
on the Keck I telescope. R. Ellis obtained two exposures of 900
seconds each on UT 2000 September 29 using the $1\farcs0$ slit with
the 600 $\rm lines~mm^{-1}$ grating that provides a resolution of $\sim$
1,000. The two exposures were taken at slightly different positions on
the slit. Data reduction followed similar procedures to those employed
for ESI. The two images were then subtracted from each other so as to
yield a sky-subtracted image. We used a NeAr+Hg,K arc lamp to
wavelength calibrate the data.
 
\subsection{HST Imaging} 

As part of an {\em HST} cycle 9 program we observed GRB~000926 at five
epochs with the Wide Field Planetary Camera 2 (WFPC2). We reported
results from the first four epochs, taken between UT~Oct 7.25 2000 and
UT~Dec 16.9 2000, in \citet{hys+01}.  In
these early epochs emission from the OT made a significant
contribution to the total measured flux from the GRB position.  We
observed at a fifth epoch, UT~May 19.63 -- 20.86 2001, in the F450W,
F606W, and F814W WFPC2 filters, and we present these results along
with a comparision to previous epochs here.

HST was pointed such that the optical transient falls on WFPC CCD\#3 
at the WFALL position. The 2200 seconds (1 orbit each) F450W images were
combined using the STSDAS task \texttt{crrej}.  The F606W and F814W images
were observed at two offsets, by +2.5, +2.5 pixels in x and
y. These images were combined and cosmic-ray rejected using the
\texttt{drizzle} technique \citep{fh97}. The drizzled images have pixels
half the area of the original WFPC data.       

\section{Metallic line absorption} 
 
\begin{figure}[]
\epsscale{0.65} 
\plotone{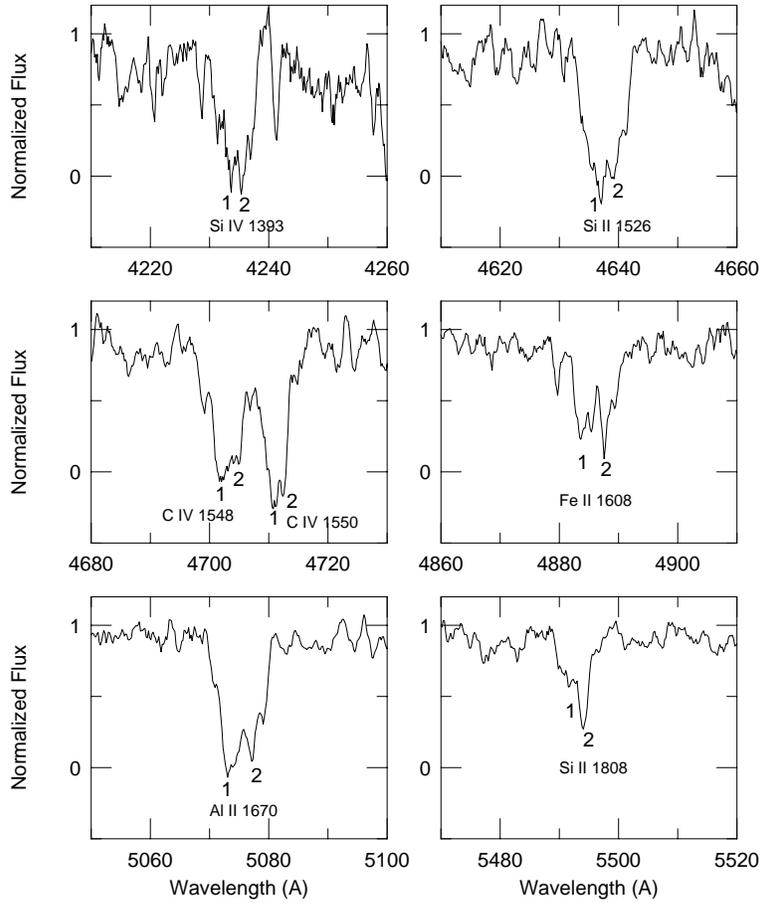} 
\caption{A zoom in of the ESI spectrum of GRB\,000926 on UT 2000 
September 29.26 showing the existence of two components with a mean 
separation of $\sim$ 4.05 $\rm \AA$ at redshifts of $\rm z_1=2.0370 
\pm 0.0011$ (indicated by 1) and $\rm z_2=2.0387 \pm 0.0011$ 
(indicated by 2). The spectrum has been smoothed by a boxcar of 7 
pixels, equivalent to 1.12 $\rm\AA$. } 
\label{fig:esispec2} 
\end{figure}

\begin{figure}[] 
\epsscale{0.65}
\plotone{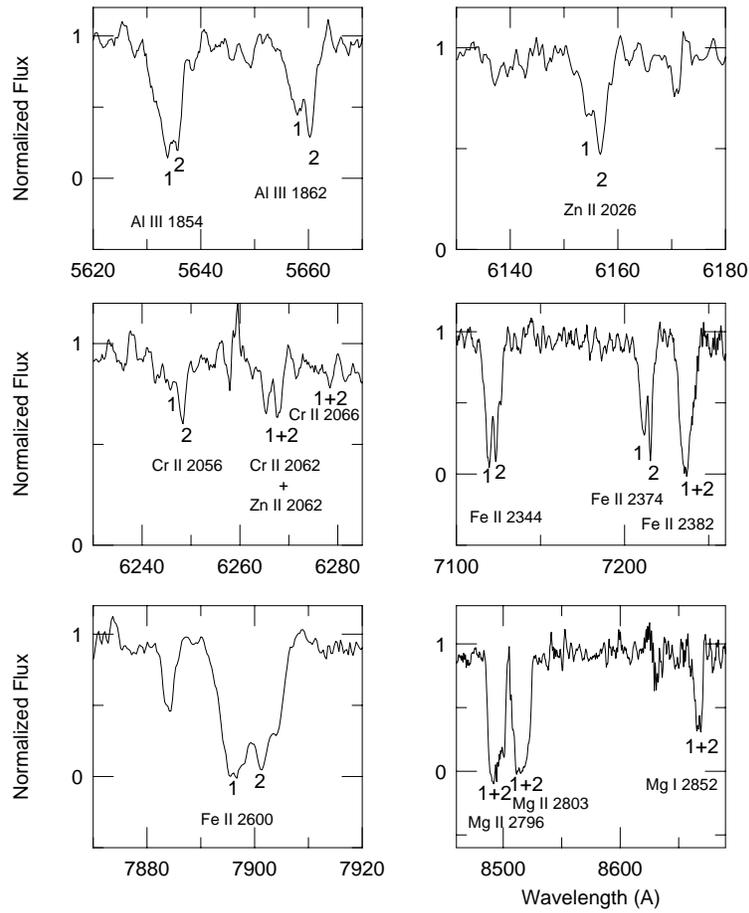}
\caption{Another set of lines in the ESI spectrum, showing the
two components of the absorption system.} 
\label{fig:esispec3} 
\end{figure}

\begin{figure}[] 
\epsscale{0.75}
\plotone{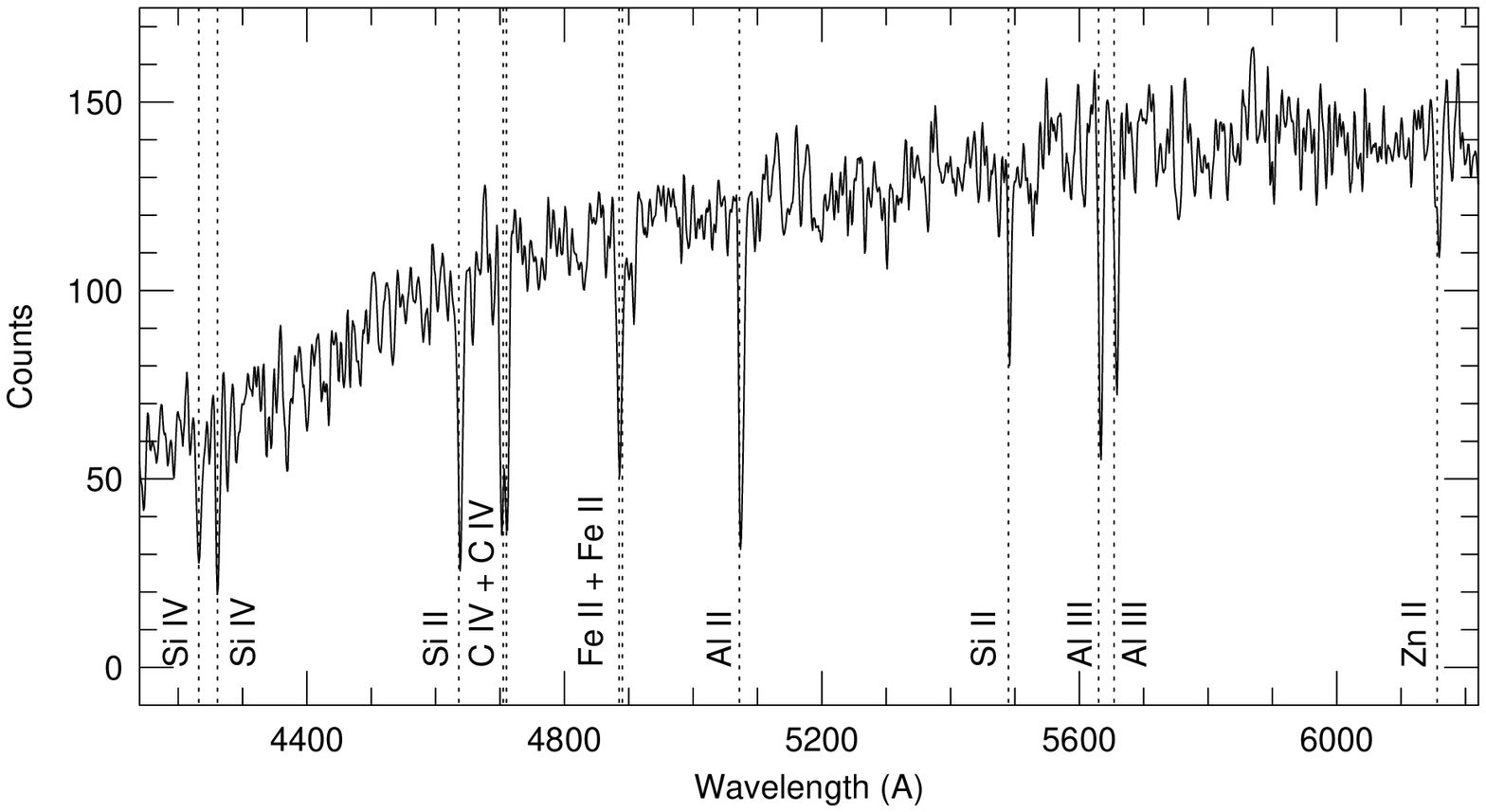} 
\caption{The LRIS spectrum of GRB\,000926 on UT 2000 September 
29**. Indicated are the line identifications corresponding to a 
redshift $z = 2.0373 \pm 0.0011$. The spectrum 
has been smoothed by $\rm\sim$3.8 \rm\AA. } 
\label{fig:lrisspec} 
\end{figure}

\citet{fmg+00} found, based on low
resolution spectra, that the optical afterglow of GRB\,000926 had two
significant absorption systems: one at $z \sim$ 1.378 and a second at
$z \sim$ 2.066. However, our higher resolution spectra show only one
system, split into two components, identified by the lines listed in
Table \ref{tab:ewesi}. The mean redshift obtained from these lines is $z
= 2.0373 \pm 0.0011$ (LRIS) and $z = 2.0379 \pm 0.0008$ (ESI),
presumably corresponding to the system identified by Fynbo {\it et
al.} as having $z \sim$ 2.066. We do not see any evidence of another
absorption system in these spectra.

\begin{deluxetable}{lccc} 
%\small 
%\footnotesize  
%\scriptsize 
%\singlespace 
%\tablewidth{504pt} 
\tablewidth{0pt} 
%\tablecolumns{6} 
\tablenum{1} \tablecaption{The ESI line identifications. 
[1] and [2] indicate measurements for the two velocity components, and
[b] is used where these cannot be separated.
Systematic and statistical (in parentheses)
errors are indicated.\label{tab:ewesi}}
\tablehead{ \colhead{line} & \colhead{$\lambda_{\rm vac}$($\rm\AA$)} & 
\colhead{z$_{abs}$} & W$\rm _{\lambda,rest}$ ($\rm\AA$) } \startdata 
Si IV 1393.76 & 4233.25 [1] & 2.03729 & 1.47 $\pm$ 0.15 (0.06) \\
Si IV 1393.76 & 4236.27 [2] & 2.03945 & 0.66 $\pm$ 0.07 (0.06) \\ 
Si II 1526.72 & 4635.80 [1] & 2.03644 & 1.41 $\pm$ 0.14 (0.02) \\ 
Si II 1526.72 & 4639.62 [2] & 2.03895 & 1.23 $\pm$ 0.12 (0.02) \\ 
C IV 1548.20 & 4701.93 [1] & 2.03703 & 2.00 $\pm$ 0.20 (0.02) \\ 
C IV 1548.20 & 4704.73 [2] & 2.03884 & 0.60 $\pm$ 0.06 (0.02) \\ 
C IV 1550.77 & 4710.86 [1] & 2.03775 & 2.24 $\pm$ 0.22 (0.02) \\ 
C IV 1550.77 & 4712.63 [2] & 2.03890 & 0.22 $\pm$ 0.02 (0.02) \\ 
Fe II 1608.46 & 4883.82 [1] & 2.03633 & 0.63 $\pm$ 0.06 (0.02) \\ 
Fe II 1608.46 & 4887.86 [2] & 2.03884 & 1.52 $\pm$ 0.15 (0.02) \\ 
Al II 1670.81 & 5073.30 [1] & 2.03643 & 1.41 $\pm$ 0.14 (0.02) \\ 
Al II 1670.81 & 5077.65 [2] & 2.03903 & 1.12 $\pm$ 0.11 (0.02) \\ 
Si II 1808.00 & 5491.10 [1] & 2.03711 & 0.36 $\pm$ 0.04 (0.01) \\ 
Si II 1808.00 & 5494.10 [2] & 2.03877 & 0.77 $\pm$ 0.08 (0.01) \\ 
Al III 1854.72 & 5633.78 [1]& 2.03754 & 1.46$\pm$ 0.15 (0.01) \\ 
Al III 1854.72 & 5635.83 [2]& 2.03864 & 0.11$\pm$ 0.01 (0.01) \\ 
Al III 1862.78 & 5657.75 [1]& 2.03726 & 0.81$\pm$ 0.08 (0.01) \\ 
Al III 1862.78 & 5660.42 [2]& 2.03869 & 0.46$\pm$ 0.05 (0.01) \\ 
Zn II 2026.14 & 6154.26 [1] & 2.03741 & 0.34 $\pm$ 0.03 (0.01) \\ 
Zn II 2026.14 & 6156.84 [2] & 2.03870 & 0.56 $\pm$ 0.06 (0.01) \\ 
Cr II 2056.25 & 6245.46 [1] & 2.03730 & 0.33 $\pm$ 0.03 (0.01) \\ 
Cr II 2056.25 & 6248.23 [2] & 2.03865 & 0.36 $\pm$ 0.04 (0.01) \\ 
Cr II 2062.23 & 6267.67 [b] & 2.03927 & 0.23 $\pm$ 0.02 (0.00) \\ 
Zn II 2062.66 & 6268.12 [b] & 2.03885 & 0.05 $\pm$ 0.00 (0.00) \\ 
Cr II 2066.16 & 6278.52 [b] & 2.03874 & 0.18$>$ W$\rm _{\lambda,rest}$ $>$0.09  \\ 
Fe II 2344.21 & 7119.02 [1] & 2.03685 & 1.77 $\pm$ 0.18 (0.02) \\ 
Fe II 2344.21 & 7123.84 [2] & 2.03891 & 1.64 $\pm$ 0.16 (0.02) \\ 
Fe II 2374.46 & 7211.09 [1] & 2.03694 & 1.75 $\pm$ 0.18 (0.01) \\ 
Fe II 2374.46 & 7215.46 [2] & 2.03878 & 0.82 $\pm$ 0.08 (0.01) \\ 
Fe II 2382.76 & 7236.64 [b] & 2.03708 & 3.07 $\pm$ 0.31 (0.02) \\ 
Fe II 2600.18 & 7896.06 [1] & 2.03673 & 1.96 $\pm$ 0.20 (0.02) \\ 
Fe II 2600.18 & 7902.28 [2] & 2.03913 & 1.69 $\pm$ 0.17 (0.02) \\ 
Mg II 2796.35 & 8490.52 [1] & 2.03629 & 2.97 $\pm$ 0.30 (0.03) \\ 
Mg II 2796.35 & 8498.62 [2] & 2.03918 & 2.38 $\pm$ 0.24 (0.03) \\ 
Mg II 2803.53 & 8511.47 [1] & 2.03598 & 2.77 $\pm$ 0.28 (0.02) \\ 
Mg II 2803.53 & 8519.58 [2] & 2.03887 & 2.64 $\pm$ 0.26 (0.02) \\ 
Mg I 2852.97 & 8665.87 [1] & 2.03749 & 1.99 $\pm$ 0.20 (0.02) \\ 
Mg I 2852.97 & 8669.25 [2] & 2.03868 & 0.44 $\pm$ 0.04 (0.02) \\ 
\enddata 
\end{deluxetable}

\begin{figure}[] 
\epsscale{0.65}
\plotone{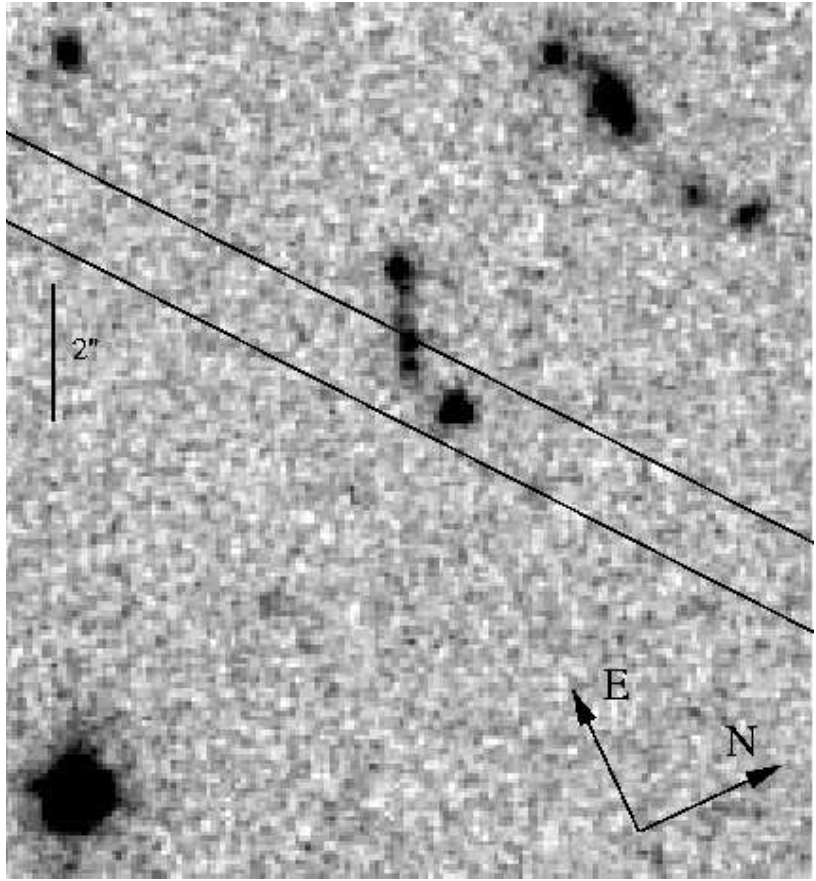}
\caption{Combined HST WFPC2 F606W image of the GRB\,000926 optical afterglow. 
Indicated is the approximate position and orientation of the ESI slit.}
\label{fig:hstesislit} 
\end{figure}

Figure \ref{fig:esispec} shows the ESI spectrum smoothed by a boxcar
of 7 pixels, equivalent to 1.12~$\rm\AA$. The absorption lines show
evidence for two components with a mean separation of $\sim$ 4.05~$\rm
\AA$ (seen in Figure \ref{fig:esispec2} and \ref{fig:esispec3}). 
Figure \ref{fig:lrisspec} shows the LRIS spectrum smoothed by
$\rm\sim$3.8~$\rm\AA$.  Table
\ref{tab:ewesi} shows the line identifications from ESI for each of
the two components, the corresponding redshift and the measured
restframe equivalent widths (EWs) $W_{\lambda, \rm rest}$. Whenever it
is not possible to separate the components, we list the equivalent
width of the blended line. Table~\ref{tab:ewlris} shows the
identifications from the LRIS spectrum, where the absorption lines
from the two systems are all blended due to the lesser resolving
power.  The errors listed in the table are the total systematic
plus statistical uncertainties, and are dominated by the 
uncertainty in the placement
of the continuum.
The statistical errors are also provided in parentheses.  We
assumed Voigt profiles for all the measured lines. We normalized the
spectra with the continuum set to unity. The derived redshifts of the
two components are $z_1=2.0370 \pm 0.0011$ and $z_2=2.0387 \pm
0.0011$, with a mean of $z = 2.0379 \pm 0.0008$.  Figure
\ref{fig:hstesislit} shows the approximate position and orientation of
the ESI slit in an HST WFPC2 F606W image of the field. Contaminating
flux from the underlying GRB host galaxy, which would cause the
equivalent widths to be underestimated, is negligible in our
spectroscopic observations (see below).

\begin{deluxetable}{lccc} 
%\small 
%\footnotesize  
%\scriptsize 
%\singlespace 
%\tablewidth{504pt} 
\tablewidth{0pt} 
%\tablecolumns{6} 
\tablenum{2} \tablecaption{The LRIS line identifications. 
Systematic and statistical (in parentheses)
errors are indicated.\label{tab:ewlris}}
\tablehead{ \colhead{line} & \colhead{$\lambda_{\rm vac}$($\rm{\AA}$)} & 
\colhead{z$_{abs}$} & W$\rm _{\lambda,rest}$ ($\rm{\AA}$) } \startdata 
 Si IV 1393.76 & 4232.52 & 2.03676 & 2.46 $\pm$ 0.25 (0.03)\\ 
Si II 1526.72 & 4638.35	& 2.03811 & 3.31 $\pm$ 0.33 (0.01)\\ 
C IV 1548.20 & 4702.54  &2.03742 & 1.88 $\pm$ 0.19 (0.02)\\ 
C IV 1550.77 & 4711.15 &   2.03794 & 1.91 $\pm$ 0.19 (0.02)\\ 
Fe II 1608.46 & 4886.10	 & 2.03775 & 2.57 $\pm$ 0.26 (0.01)\\ 
Al II 1670.81 & 5075.81	 & 2.03613 & 2.47 $\pm$ 0.25 (0.03)\\ 
Si II 1808.00 & 5491.96	 & 2.03759 & 1.39 $\pm$ 0.14 (0.02)\\ 
Al III 1854.72 & 5633.60 & 2.03744 & 2.06 $\pm$ 0.21 (0.02)\\ 
Al III 1862.78 & 5655.15 & 2.03586 & 1.95 $\pm$ 0.19 (0.01)\\ 
Zn II 2026.14 & 6155.67	& 2.03813 & 1.20 $\pm$ 0.12 (0.01)\\ 
\enddata 
\end{deluxetable} 
 
A comparison of the equivalent widths measured from the LRIS spectrum 
with those from the ESI spectrum (summing the two components) shows 
generally good agreement, apart from a small tendency for larger 
equivalent widths in the LRIS spectrum. This can be understood as the 
effect of not being able to resolve the two systems in the lower 
resolution LRIS data.  The ESI measurements are therefore more  
accurate. 
 
\section{The GRB~000926 Host Galaxy and Offset}

\begin{figure}[] 
\epsscale{0.65}
\plotone{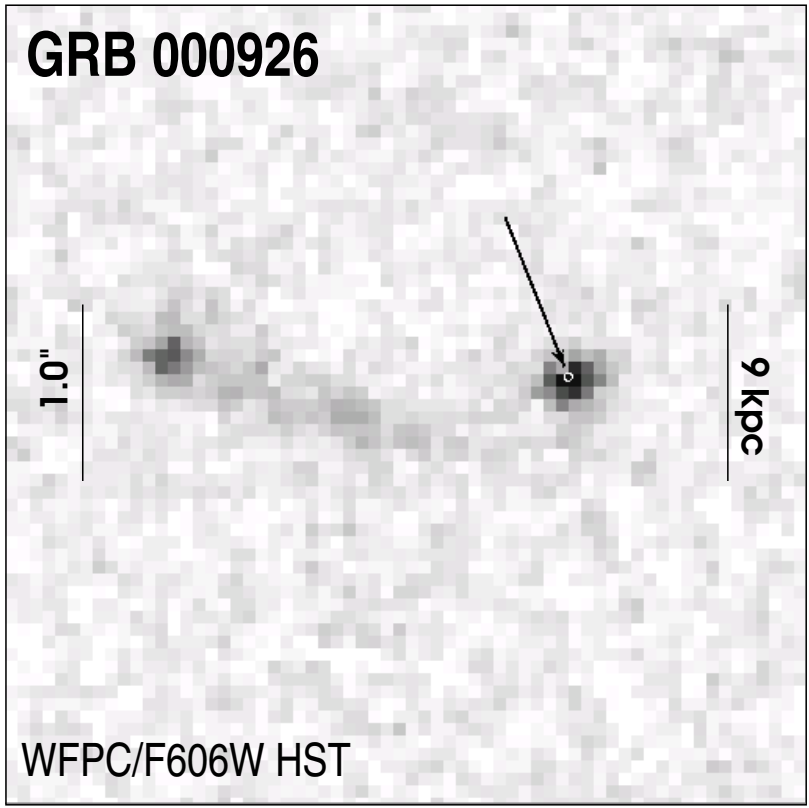} 
\caption{Combined HST WFPC2 F606W image of a 
$4.5^{\prime\prime} \times 4.5^{\prime\prime}$ region surrounding
GRB\,000926, taken on UT May 19.72, 2001, 238 days after the event.
At this epoch the emission from the GRB position is dominated by the
host, seen to be compact and relatively bright (24.83~mag in $R$). The
white ellipse is the 9-sigma contour position of the OT.}
\label{fig:hstimage}
\end{figure}  

\begin{figure}[] 
\epsscale{0.65}
\plotone{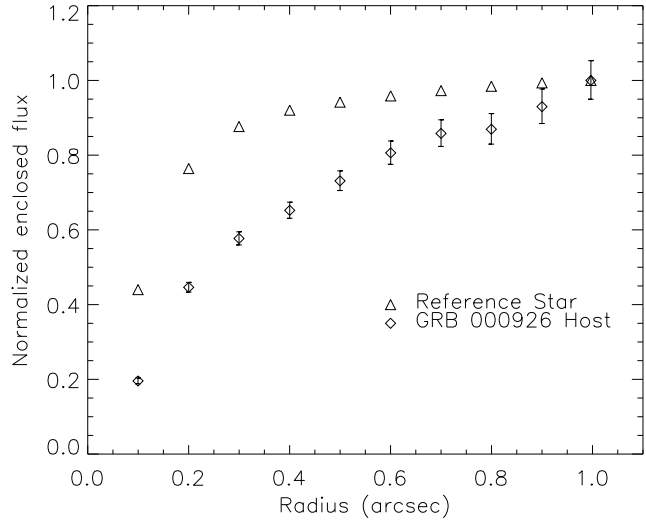} 
\caption{Emission from the region uderlying the GRB~000926 optical 
transient as a
function of extraction radius (diamonds) compared to a reference star
(triangles).  The vertical scale is magnitudes normalized to unity at
0.1\arcsec\ radius.  This clearly shows the know of emission to be
extended compared to the instrument point spread function.}
\label{fig-hostext}
\end{figure}  

\begin{figure}[] 
\plotone{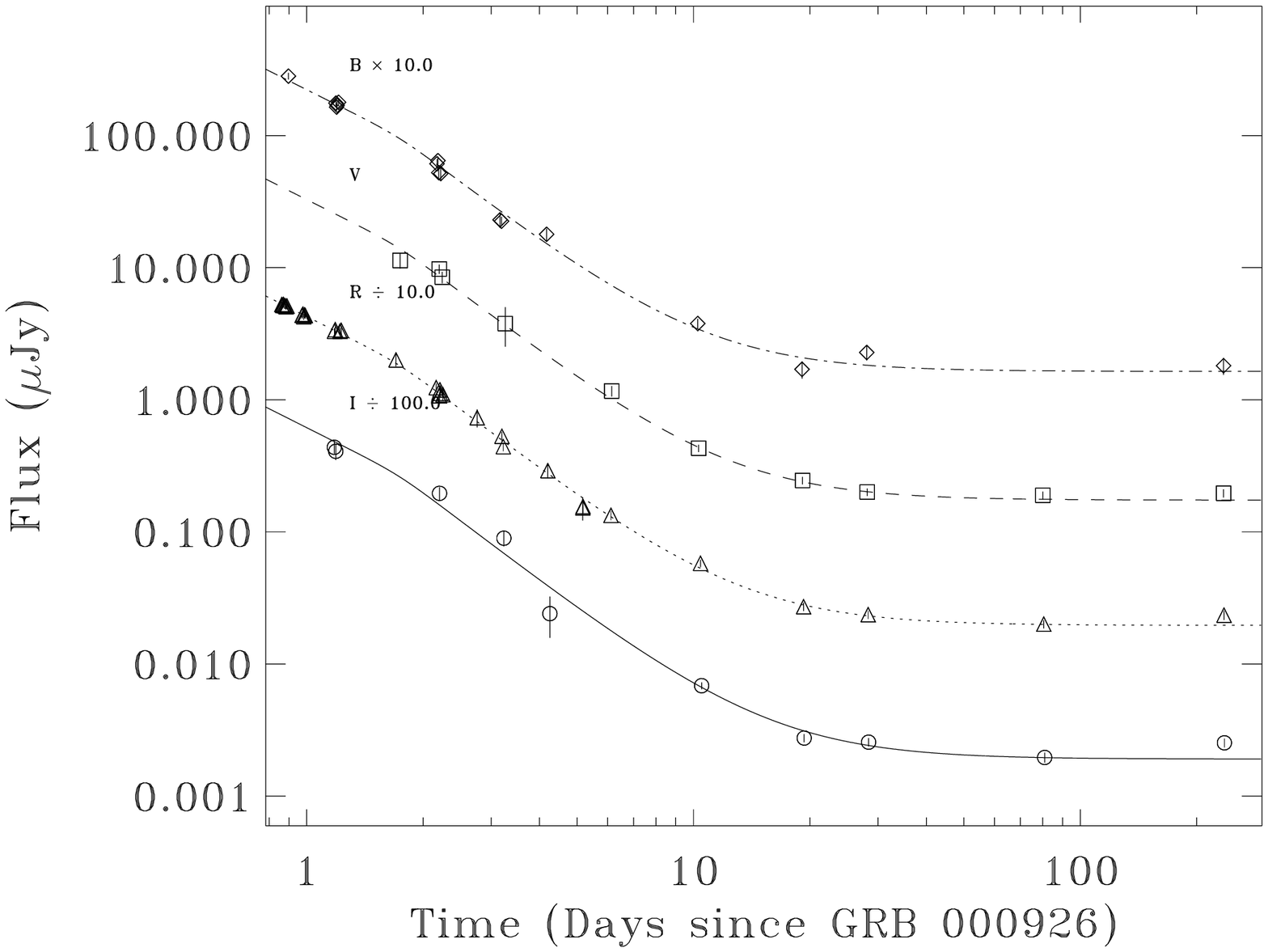} 
\caption{Optical lightcurve of GRB~000926. The data are from
Price {\em et al.} (2001) except for the last epoch, taken with HST
WFPC 2 on UT May 19.26 -- 20.86. The solid line shows a model fit
to the afterglow emission with a constant host contribution added.}
\label{fig:lc} 
\end{figure}

In the final epoch HST images the flux from the region is due entirely
to the host galaxy. Figure~\ref{fig:hstimage} shows the WFPC2 F606W
image from UT~May~19.72 (238 days after the
GRB). Table~\ref{tab:hstmags} provides the magnitude of the emission
extracted from a 0.2\arcsec\ radius region surrounding the position of
the optical transient.  These magnitudes are aperture corrected to
0.5\arcsec\ to match published calibrations \citep{hbc+95}. We then
calibrated the WFPC2 data using the zeropoints from \citet{hbc+95}
(Table 10 page 1086). Table~\ref{tab:hstmags} shows the resulting
SMAGs. We then transformed to the Johnson Cousins system using
published color transformations (see Table 10 from Holtzman et
al. 1995). The transformed magnitudes are provided in Table
\ref{tab:BVRI}. We note that the quoted errors are statistical only.

\begin{deluxetable}{cccc}
\tablewidth{0pt} 
\tablenum{3} \tablecaption{HST WFPC2 observations of the GRB~000926 
host emission in 2-pixel (0.25$\arcsec$) radius region 
centered on the OT position.\label{tab:hstmags}}
\tablehead{ \colhead{epoch} & \colhead{filter} &
\colhead{$t_{obs}$ (s)} & \colhead{mag }}
\startdata
19.63  &    F450W  &   2200 & $25.80 \pm 0.14$    \\
19.72  &    F606W  &  4400  &  $25.47 \pm 0.05$   \\
19.86  &    F814W  &  4400 &  $24.97 \pm 0.06$   \\
\enddata
\end{deluxetable}

\begin{deluxetable}{ccc}
\tablewidth{0pt} 
\tablenum{4} \tablecaption{$BVRI$ magnitudes for the
emission extracted from HST images in
 2-pixel (0.25$\arcsec$)  
radius region centered on the OT.\label{tab:BVRI}}
\tablehead{ \colhead{band} &
\colhead{mag (2 pix)}  & \colhead{mag (5 pix)}}
\startdata
B & $25.86 \pm 0.14$ & $25.49 \pm 0.33$   \\
V  &  $25.66 \pm 0.05$ & $25.08 \pm 0.06$  \\
R &  $25.29 \pm 0.06$ & $24.83 \pm 0.07$  \\
I &  $24.95 \pm 0.06$ & $24.59 \pm 0.01$  \\    
\enddata
\end{deluxetable}
 
To determine the extent of the knot of emission underlying the optical
transient we compared the radial profile to that of a standard star.
Figure~\ref{fig-hostext} shows the flux extracted in increasing
apertures for the host emission (diamonds) compared to the reference
star. The scale is normalized to unity for a 1\arcsec\ radius
aperture.  The emission is clearly extended compared to the instrument
point spread function.

To determine the precise location of the OT in the bright knot of
emission (which we take to be the host galaxy), we compared the OT
centroid taken from the first epoch HST WFPC F606W image (when the OT
emission was dominant; see Figure~\ref{fig:lc}) and compared it to the
centroid of the knot seen in the late-epoch image.  To register the
first-epoch to the final image, we
drizzled both images independently increasing the drizzled pixel size
by a factor of 2 in area ({\tt PIXFRAC = 0.8}). We then registered the
final stacked images from the two epochs using IRAF/CROSSCOR and
IRAF/SHIFTFIND restricting the cross-correlation to the region of
overlap between the two epochs on WFPC chip \#3.  Using IRAF/CENTER
and OFILTER centering algorithm, we computed the centers of the OT and
the host.  The result is a statistical offset of $23.9
\pm 4.2$ mas E, $21.2 \pm 4.1$ mas N (host$\rightarrow$OT) or $r =
31.9 \pm 4.1$ (note that the uncertainty of 4.1 mas is not the normal
Gaussian r.m.s; see \citet{bkd01} for further details).  As a demonstration of
the good registration of the images, we note that the bright
knot of emission located 2\arcsec .57 E and 0\arcsec .15 of the host
had a statistically negligible offset of (epoch 5$\rightarrow$epoch 1)
$11.7 \pm 6.1$ mas W, $8.4 \pm 6.0$ mas S between images.  The white
oval in Figure~\ref{fig:hstimage} shows the position of the OT
compared to the host.

%In the late-epoch image the majority of the flux from the OT position
%is from the host galaxy.  Extrapolating the transient lightcurve,
%the OT should be have an R magnitude of 33.6, much fainter than the
%25.29~mag measured from a 2-pixel radius surrounding the OT location.
% Finally, the lightcurve for the
%emission extracted from the OT position flattens after $\sim 30$ days
%(Figure~\ref{fig:lc}), indicating that after this the dominant
%emission is from a constant source.

In addition to the bright, compact knot, Figure~\ref{fig:hstimage}
also shows an arc of emission extending East from the OT location.
This arc is separated by a projected distance of $\sim$4~kpc from the
compact knot.  Ground-based narrow-band Ly-alpha imaging indicates
that the extended emission also originates from $z = 2.04$
\citep{fyn01}, and therefore may be associated with the host.  The
two bright knots, separated by $\sim 20$~kpc are possibly the nuclei
of interacting systems. Comparing the emission extracted from a
5-pixel radius surrounding the OT and the Eastern knot in the F814W
and F606W WFPC filters, the knot coincident with the OT appears to be
somewhat redder at the 3.5-$\sigma$ level.

\section{Discussion} 

Two notable features of the GRB~000926 optical transient are the small
offset from the center of the host galaxy, and the large equivalent
widths of the absorption lines.  At the
redshift of GRB 000926, the OT offset amounts to \hbox{$287
\pm 37$ pc} projected on the sky.  Only 2 out of 15 other GRBs have
smaller measured offsets from their hosts (GRB 000418 and GRB 970508;
\citet{bkd01}).  The absorption line equivalent widths are 
large compared to those seen in other GRBs.  The Mg~II absorption,
with $W_{\lambda}$(Mg II 2796.3) = $5.4 \pm 0.5 \,\rm\AA$ and
$W_{\lambda}$(Mg II 2803.5) = $5.4
\pm 0.5 \,\rm\AA$ (summing components 1 and 2), is by far the largest 
measured in a GRB afterglow to date.  The Mg II is highly saturated;
the doublet ratio $W_{\lambda}$(Mg II 2796.3)/$W_{\lambda}$(Mg II
2803.5) is close to unity, and the ratio of Mg I to Mg II,
$W_{\lambda}$(Mg I 2853.0)/$W_{\lambda}$(Mg II 2796.3) = 0.45 $\pm$
0.06 is very high. In quasar absorption line systems this ratio is
typically $\leq$ 0.15 \citep{ss92}.  Such a high ratio was also noted
in GRB\,970508 \citep{mdk+97}.

Combined with the small offset, the high EWs indicate that the
GRB~000926 OT lies close to the center of the compact host.  The small
offset alone means the projected distance is small, but the measured
EWs imply the line of sight subtends a significant column of material,
in excess of that expected for a typical galactic halo. Damped Lyman
alpha systems are believed to be normal foreground galaxies, and the
lines of sight to the backlighting quasars most often subtend the halo
of the absorber. The Mg~II features seen in damped Lyman alpha systems
at similar redshift typically have $W_{\lambda}$(Mg II 2796.3) between
0 and 3 $\rm\AA$, falling off steeply towards high $W_{\lambda}$
\citep{ss92}. The sample of \citet{ss92} contains no system with
$W_{\lambda}$(Mg II 2796.3)$ > 3\rm\AA$. Our measurement of
$W_{\lambda}$(Mg II 2796.3) = $5.4 \pm 0.5 \,\rm\AA$ is therefore
exceptionally high.  The large relative values we measure for
GRB~000926 therefore imply the OT was embedded in the galactic disk.

Absorption redshifts provide only lower limits, however in this case,
as argued above, the absorption is almost certainly due to the host
galaxy.  Assuming a $\Omega_{\rm m} = 0.3$, $\Omega_{\lambda}= 0.7$,
$H_0$ = 65 km s$^{-1}$ Mpc$^{-1}$ cosmology the measured redshift and
the k-corrected 20 - 2000 keV fluence corresponds to an isotropic
energy release of (2.97 $\pm$ 0.10) $\times 10^{53}$ erg, which
reduces to 3 $\times 10^{51}$ erg when corrected for collimation of
the outflow \citep{hys+01}.
 
The two components seen in the ESI absorption lines are likely due to
individual systems in the host galaxy.  The velocity separation of 168
km~s$^{-1}$ is typical for relative motions between clouds in a
galaxy.  Adopting the assumption that the GRB lies in one of these
clouds near the galactic center, and the second cloud is at the
visible half light radius so that the measured velocity is
representative of galactic rotation at this distance, we can estimate
the host galaxy mass to be $M_{host} \sim v^2 r_{1/2} G^{-1} =
10^{10}~M_{\odot}$.

In principle, absorption spectroscopy can be used to measure metallicity
in distant galaxies, and it is interesting to ask how metal abundances
in GRB hosts compare to other systems at similar redshift.  
Unfortunately our spectra do not extend far enough into the blue to 
measure Ly$\alpha$, so we cannot make any inference about the 
metallicity of the host galaxy or GRB site.  Because the degree of metal 
enrichment in galaxies at the same epoch can vary by more than two 
orders of magnitude \citep{psh+94}, we also cannot estimate the hydrogen 
column from the EWs of the metal absorption lines. 
 
The relative abundances of various metals can, however, be used to
make some inferences about the physical properties of the absorbing
medium.  In particular, the relative abundances of zinc and chromium,
as measured from the Zn~II (2026.15,2062.66) and Cr~II (2056.25,
2062.23, 2066.16) lines provide a qualitative estimate of the dust to
gas ratio in the absorbing clouds.  In a study of these lines in
damped Lyman-alpha systems, Pettini {\em et al.} (1990) noted that the
Zn~II doublet provides a good measurement of Zn in the gas phase of
the ISM, and further that Zn is not readily incorporated into dust
grains.  Interstellar Cr, on the other hand, is one of the most
heavily depleted elements (see Pettini {\em et al.} (1990) for more
detailed discussion).
 
We detect Zn~II 2026 and Cr~II 2056 with high statistical significance 
in a region of the spectrum where we can accurately measure the 
continuum, and these EWs are listed in Table~\ref{tab:ewesi}.  We cannot 
resolve the Cr~II 2062.23 and Zn~II 2062.66 lines, but we measure the 
combined redshift-corrected equivalent width of $W = 0.23 \pm 0.02 
(0.01) \rm\AA$ ($z = 2.03927$) and $W = 0.05 \pm 0.00 (0.01) \rm\AA$ ($z = 
2.03885$) for the blend.  The Cr~II 2066 line is in a region 
where it is very difficult to measure the continuum.  Based 
on upper and lower limits on the continuum, we get $0.18 > W > 0.09 
\rm\AA$ ($z = 2.03874$).  From the line ratios, we can infer that the 
Zn~II and Cr~II lines are likely not saturated, and we can therefore 
use the EWs to determine the relative abundances of Zn and Cr. 

The Cr/Zn ratio we measure for the absorbers in the GRB~000926 host
indicates the host galaxy is depleted in dust relative to the local
ISM. The ratio [Cr/Zn] $= log[N(Cr)/N(Zn)] - log[N(Cr)/N(Zn)]_{\odot}$
for both components of the absorption system is significantly larger
than that measured locally. Using the oscillator strengths from
Pettini {\em et al.}  (1994), we obtain [Cr/Zn]$ = -0.92 \pm 0.11$ and
$-1.1 \pm 0.13$ for the two components of the GRB absorber, indicating
the gas phase abundance of Cr is almost a factor 10 higher than in the
local ISM.  [Cr/Zn] measured for the GRB~000926 clouds is in fact
quite similar to the range $-1.8$ to $ -2$ that Pettini {\em et al.}
(1994) obtain for damped Ly$\alpha$ systems between $2 < z < 3$. For
the damped Ly$\alpha$ systems, Pettini {\em et al.} (1994) conclude
that this is due to the depletion of dust in high redshift galaxies.
A similar qualitative statement applies to the GRB~000926 host, again
assuming that the absorption systems are local to the host.
 
The fact that $[Cr/Zn]$ is similar in the absorption spectra of the
two clouds indicates the dust depeletion is a characteristic of the
galaxy as a whole, rather than of the immediate GRB environment.  It
has been suggested that GRBs reside in gas-rich regions, and the dust
in this gas is destroyed by heating and sublimation of the grains by
the X-ray and UV flux \citep{wd00} (effective to about 10 pc) or/and
by ejection of electrons by X-rays and subsequent shattering of the
grains by the repulsive force of the net charge left \citep{fkr01}
(effective out to about 100 pc).  This suggestion has been used to
explain the observation that in a sample of events, the optical extinction
towards GRBs is lower than one would expect from measured X-ray column
densities \citep{gw00}.  The postulated destruction of dust takes
place locally to the GRB, however our measurements find dust to be
similarly depleted in clouds that must be separated by much more than
10~pc.  We suggest therefore that the observed X-ray columns result
from absorption in the disks of the host galaxies, and that the low
extinctions seen in many OT optical spectra are due to the lack of
dust in galaxies at these redshifts.

\section{Conclusions} 
  
High-resolution absorption spectroscopy of GRB~000926 strengthens the
case that GRBs explode in the disks of galaxies.  The small offset
from the center of the host, combined with the evidence for a
significant column depth of metals indicates the explosion occurred
near the center of a galaxy at redshift $z = 2.038$.  The two
absorption systems seen in the intermediate-resolution spectra,
separated by $v = 168$~km~s$^{-1}$, are most easily interpreted as due
to separate clouds within the host galaxy.  The similarity in the
properties (e.g. ratios of metals) between the two clouds indicates
that GRBs do not significantly alter their immediate environment.
Rather, differences between dust to gas ratios in GRB host galaxies to
the local sample are most likely due to evolutionary effects.
 
The ability to measure the EWs of a significant number of metal lines
in the GRB~000926 spectrum suggests that in the future high-resolution
studies of optical transients will provide a means of probing the
metallicity and dust content of GRB hosts.  For GRB~000926 we were
unable to measure the Ly$\alpha$ line, so we cannot directly infer the
metal enrichment.  This should, however, be possible in future events.
From the ratio of Cr/Zn it appears that, like other galaxies at
similar redshift, the dust to gas ratio is reduced in the GRB~000926
host compared to local values.  With a larger sample it should be
possible to determine if there are any significant differences in
metal enrichment and dust depletion in GRB host galaxies relative to
other distant galaxies.
 
\clearpage

%\bibliographystyle{apj}  
%\bibliography{journals_apj,grbrefs} 

\end{document}